\documentclass[prb,aps,twocolumn]{revtex4}
\usepackage{epsfig,latexsym,amssymb,amsmath,amsbsy,graphics,graphicx}

\begin{document}
\def \beq{\begin{equation}}
\def \eeq{\end{equation}}
\def \beqarr{\begin{eqnarray}}
\def \eeqarr{\end{eqnarray}}

\title{Thermodynamics of strongly disordered spin ladders}

\author{Eddy Yusuf and Kun Yang}
\affiliation{
National High Magnetic Field Laboratory and Department of Physics,
Florida State University, Tallahassee, Florida 32306
}

\begin{abstract}
We study antiferromagnetic two-leg spin-1/2 ladders with strong bond randomness,
using the real space
renormalization group method. We find the low-temperature spin susceptibility of
the system follows non-universal power laws, and the ground state
spin-spin correlation is short-ranged. Our results suggest that there is no 
phase transition when the bond randomness increases from zero; for strong
enough randomness the system is in a Griffith region with divergent
spin susceptibility and short-range spin-spin correlation.
\end{abstract}
\maketitle

\section{Introduction}

One-dimensional (1D)
spin systems are of interest to physicists since the early days
of quantum mechanics.\cite{bethe} Considerable effort has been devoted to the
theoretical 
study of antiferromagnetic (AF) spin chains, where some of the very few exact
solutions of interacting Hamiltonians in physics were obtained,\cite{bethe,aklt}
and remarkably
rich low-energy physics were uncovered using various non-perturbative 
methods.\cite{haldane}
Interest in these systems were also enhanced by the recent experimental 
realization of such model systems,\cite{expt}
due to technological advances. More recently,
considerable attention has focused on another class of 1D spin systems, 
namely AF spin ladders.\cite{dagotto} These systems are made of two or more 
coupled spin chains. The physics of such systems are closely related to, but
even
richer than the spin chain systems. Further motivation for study of these systems
comes from the similarity in structure
between these systems and undoped cuprates, and the
discovery of superconductivity in them once charge carriers are introduced 
via doping. 

The ubiquitous randomness is known to have particularly strong effects in 
low-dimensional systems. Recently, there has been rather extensive theoretical
studies of effects of disorder in spin chains. Most of these studies are based
on the celebrated real space renormalization group (RSRG) method introduced
by Ma, Dasgupta and Hu (MDH) in the study of AF spin-1/2 chain with bond 
randomness,\cite{mdh} and Bhatt and Lee in the study of magnetic properties of
doped semiconductors.\cite{bl} This method was elaborated and extended in 
great detail by Fisher,\cite{fisher1} and applied (often with nontrivial 
extensions) by a number of other authors to
various disordered spin chain 
models.\cite{fisher2,westerberg,hybg,boechat,hy,monthus,yb,damle}
A variety of disorder-dominated phases have been found, whose low-energy
physics is qualitatively different from their disorder free counterparts.
While the quantitative accuracy of the RSRG relies on the presence of {\em strong} randomness, it has been shown\cite{fisher1} that even if the strength of
randomness is weak, it tends to grow as the RSRG proceeds to lower and lower
energy scales, thus giving qualitatively correct (and sometimes asymptotically
exact) low-energy behavior. Indeed, many predictions of RSRG have been 
confirmed by complementary analytical and numerical studies using other methods.

Comparatively speaking, relatively few studies have focused on effects of 
randomness on spin ladders. Effects of doping by non-magnetic impurities 
(or site dilution) have
been studied using quantum Monte Carlo\cite{iino,miyazaki} 
and mapping to Dirac fermions
with random mass.\cite{steiner,gogolin} The stability of the pure ladders 
against various types of {\em weak} randomness has been studied by Orignac
and Giamachi\cite{og} using bosonization.
In the present paper we study a two-leg AF spin-1/2
ladder with {\em strong} bond randomness, using the RSRG. 
We believe our work is complementary to the previous studies, as the effects 
of bond randomness and site dilution are quite different, and the RSRG is
particularly suitable for studies of systems with {\em strong} randomness.

While the present work was being completed, a preprint\cite{melin} 
appeared on the 
cond-mat archive, in which the authors used the RSRG as well as the density
matrix renormalization group to study various disordered ladder models. While
our work certainly overlaps with theirs, there exist two major differences.
(i) Ref. \onlinecite{melin} focuses mainly on the distribution of the gap
separating the ground and first excited states in finite clusters, while we
study mainly thermodynamic properties and spin-spin correlation functions.
(ii) Ref. \onlinecite{melin} has studied finite-size ladders with length up to
512. In our work we have studied ladders with length up to 20,000, nearly a
factor of 40 bigger. The larger size is crucial to us for 
obtaining low-temperature, large distance behavior of the thermodynamic
quantities and spin-spin correlation functions respectively. We will compare
our results with those of Ref. \onlinecite{melin} and previous studies 
whenever appropriate.

Our main results are summarized as follows. We find the thermodynamics of the
two-leg spin ladder remains {\em non-universal}, and the spin-spin correlation
remains short-ranged, even in the strong-randomness limit. This is very 
different from the random AF spin-1/2 or spin-1 chain, where weak 
(for spin-1/2\cite{doty,fisher1}) or sufficiently strong (for spin-1\cite{hy,monthus}) randomness drive the system into the Random Singlet (RS) phase with
universal thermodynamics and power-law spin-spin correlation. For sufficiently
strong randomness, the spin susceptibility of the ladders exhibits power-law
divergence as the temperature $T\rightarrow 0: \chi(T)\sim T^{-\beta}$, where
$\beta$ varies continuously with randomness strength. Combined with short-range
spin-spin correlation, this is 
characteristic of quantum Griffith behavior. Such behavior appears to persist
even when the interchain coupling of the ladder is made very weak, suggesting
weak interchain coupling immediately destabilizes the RS phase that controls
the low-energy, long-distance physics of the decoupled chains.

The remainder of the paper is arranged as follows. In section II we introduce
the model Hamiltonian we study, briefly review the RSRG method and its 
application to random spin chains, and discuss the necessary extensions we 
need to make in order to apply it to the ladder system. In section III we 
present our numerical results. In section IV we discuss the implications of
our results, make contact with related theoretical and experimental work, and
state our conclusions.

\section{Model and The Renormalization-Group Scheme}

Consider an antiferromagnetic nearest-neighbor Heisenberg spin-1/2 ladder. 
The Hamiltonian for a two-leg ladder is given by:
\begin{equation}
\label{ladder}
H = \sum_{i=1,}^{N-1}\sum_{j=1,2} J_{i,j} \boldsymbol{S}_{i,j}\cdot\boldsymbol{S}_{i+1,j} + \sum_{i=1}^{N} K_{i} \boldsymbol{S}_{i,1}\cdot\boldsymbol{S}_{i,2},
\end{equation}
where $N$ is the number of spins on a single chain, $\boldsymbol{S}_{i,j}$ 
is a spin-1/2 operator, and the positive coupling constants 
$J_{i,j}$ (couplings along the chains, or legs of the ladder) and $K_i$ 
(couplings between the chains, or along the rungs of the ladder) are 
distributed randomly according to some probability 
distributions $P_{\parallel}(J_{i,j})$ and $P_{\perp}(K_i)$. 

\begin{figure*}
\includegraphics[scale=0.6]{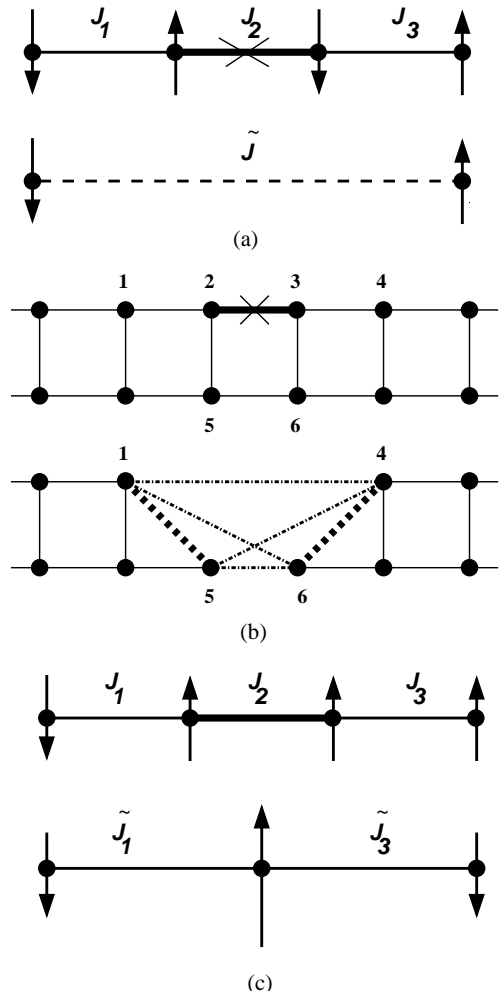}
\caption{(a) The renormalization scheme for a four-spin problem with AF 
couplings, as encountered in random AF spin-1/2 chains. Here the strongest 
bond $J_2$ is decimated, together with the neighboring bonds $J_1$ and $J_3$, 
yielding an effective interaction $\tilde{J}$ between what were the third-nearest 
neighbors. (b) Schematic diagram for decimation in ladder. The dashed lines are the renormalized couplings. The thick dashed lines are the ferromagnetic couplings generated in the decimation process. (c) The renormalization scheme for four spin problem where the strongest bond is ferromagnetic. The two spins connected by F coupling forms an effective spin object having renormalized interactions with its neighbors.}
\label{fig:rg}
\end{figure*}

In this work we use the real space renormalization group (RSRG)
method\cite{mdh} to study the
Hamiltonian (1).
We begin by briefly reviewing its application to the random AF spin-1/2  chains to
illustrate the basic ideas behind it. In this approach one identifies the 
strongest coupling of the system, say, $J_2$ that couples spins 2 and 3, and
the two neighboring spins that are coupled to this strongly coupled pair.
The Hamiltonian of this four-spin cluster is given by 
\begin{equation}
\label{h1}
H = H_0 + H_I,
\end{equation}
where
\begin{eqnarray}
\label{h2}
H_0 &=& J_{2} \boldsymbol{S}_2\cdot\boldsymbol{S}_{3}\nonumber,\\
H_I &=& J_1 \boldsymbol{S}_1\cdot\boldsymbol{S}_{2} 
+ J_3 \boldsymbol{S}_3\cdot\boldsymbol{S}_{4}.
\end{eqnarray}
In the presence of {\em strong} randomness,
$J_2$, being the strongest coupling in the system, is likely to be {\em much} 
stronger than other two couplings $J_1$ and $J_3$. In this case to a very good
approximation spins 2 and 3
form a singlet pair in the low-energy states of the entire system and become 
inert. The weak perturbation $H_I$ to this
pair induces virtual transitions to the excited (or triplet) states of $H_0$; 
the main effect of such virtual transitions is to induce an {\em effective}
coupling between spins 1 and 4 of the form:
\begin{equation}
H_{eff} =  \tilde{J} \boldsymbol{S}_1\cdot \boldsymbol{S}_4;
\end{equation}
to second order in $H_I$, $\tilde{J}$ is given by :
\begin{equation}
\label{recursion}
\tilde{J} = \frac{J_{1} J_{3}}{2 J_{2}} > 0.
\end{equation}
In essence, the RSRG procedure replaces the four spin cluster by spins 1 and 4,
which are the active degrees of freedom at low energies, with  
a new effective AF bond $\tilde{J}$ that couples them.
$\tilde{J}$ is typically
much weaker than the original bonds ($J_1, J_2$ and $J_3$)
so the bond distribution broadens and 
the energy scale is lowered. The decimation does not change the lattice 
structure of the chain, as after spins 2 and 3 are decimated, spins 1 and 4 
become nearest neighbors, thus the new effective Hamiltonian still describes an
AF spin-1/2 chain with nearest-neighbor interactions, and this procedure can be
repeated. 
The renormalization scheme is depicted in Fig. \ref{fig:rg}a.

When we apply this technique to the ladder systems, some new features not 
encountered before appear, and corresponding modifications to the spin 
decimation procedure described above need to be made. Firstly, the structure 
of the ladder gets distorted as soon as RSRG is applied, in contrast to the 
chain case. This situation requires us to keep track of the structure of the 
system as we decimate the spins and bonds. 
Secondly, when one decimates strongly 
coupled spin pairs, both AF and {\em ferromagnetic} (F) effective bonds are 
generated; these F bonds can lead to effective spins higher than spin-1/2 at low
energies. The initial renormalization step for the ladder is illustrated in 
Fig. \ref{fig:rg}b, from which we can see how the ladder structure is 
distorted and ferromagnetic interactions are generated. 
These generated F bonds are much weaker than the original bonds 
that get decimated.  
However as we 
move on, more and more spins get decimated and the energy scale is lowered 
so at some point the generated ferromagnetic bonds can become important 
because it might be the strongest bond in the system. This spin pair, instead 
of forming a singlet, form a triplet or an effective spin-1 object at 
low-energy. Clearly larger spins can also be generated by RG as energy scale
gets lowered. This situation is illustrated in Fig. \ref{fig:rg}c. 
In short, we need to keep track of both the lattice structure, and size of the
spins, together with the coupling constants (which can now be either AF or F)
in our RG procedure.

Now we turn the discussion to some technical details involved 
in the application of RG scheme to ladder. First consider two spins connected by a strong AF bond. These two spins are also coupled to the other 4 spins as shown in Fig. \ref{fig:rg}b. We will make a slight change of notation for our discussion here just for simplicity. We label the spins participating in the process by number 1 - 6. The Hamiltonian for the 6-spin cluster is given by
\begin{equation}
H = H_0 + H_I,
\end{equation}
where
\begin{eqnarray}
H_0 &=& J_{23} \boldsymbol{S}_2 \cdot \boldsymbol{S}_3,\nonumber\\
H_I &=& J_{12} \boldsymbol{S}_1\cdot\boldsymbol{S}_{2} + J_{34} \boldsymbol{S}_3\cdot\boldsymbol{S}_{4} + J_{25} \boldsymbol{S}_2\cdot\boldsymbol{S}_{5} + J_{36} \boldsymbol{S}_3\cdot\boldsymbol{S}_{6},\nonumber\\
\end{eqnarray}
where $J_{ij}$ is the bond between $\boldsymbol{S}_i$ and $\boldsymbol{S}_j$. 
This 6-spin problem is quite complicated to solve but it can be simplified using the fact that we can treat $H_I$ as a perturbation to $H_0$, especially when
the randomness is strong. It is easy to see that to second-order, $H_I$ only 
generates {\em pairwise} interaction among the spins. It is thus only necessary
to include a pair of spins coupled to the two spins connected by the strongest 
bond, when we consider the effective interaction between them.  
This fact simplifies the calculation as we can now reduce a 6-spin cluster 
problem to six 4-spin clusters\cite{note1}
which can be classified into three different 
types of 4-spin clusters as represented by: spin 1, 2, 3, and 4 (1234), (1235),
and (2356). The Hamiltonians for these clusters are given by :
\begin{eqnarray}
\label{h3}
H_1 &=& J_{23} \boldsymbol{S}_2\cdot\boldsymbol{S}_{3} + J_{12} \boldsymbol{S}_1\cdot\boldsymbol{S}_{2} + J_{34} \boldsymbol{S}_3\cdot\boldsymbol{S}_{4},\nonumber\\
H_2 &=& J_{23} \boldsymbol{S}_2\cdot\boldsymbol{S}_{3} + J_{12} \boldsymbol{S}_1\cdot\boldsymbol{S}_{2} + J_{25} \boldsymbol{S}_2\cdot\boldsymbol{S}_{5},\nonumber\\
H_3 &=& J_{23} \boldsymbol{S}_2\cdot\boldsymbol{S}_{3} + J_{25} \boldsymbol{S}_2\cdot\boldsymbol{S}_{5} + J_{36} \boldsymbol{S}_3\cdot\boldsymbol{S}_{6}.
\end{eqnarray}
$H_1$ has the same form as Eqs.(\ref{h1}) and (\ref{h2}) which lead
to the recursion relation Eq.(\ref{recursion}). 
Cluster 1235, given by the Hamiltonian $H_2$, is a new cluster structure 
not encountered in the chain case. Second order perturbation calculation shows 
that there is a new negative effective interaction between $\boldsymbol{S}_{1}$ and $\boldsymbol{S}_{5}$ given by :
\begin{equation}
\label{recursion2}
\tilde{J}_{15} = - \frac{J_{12} J_{25}}{2 J_{23}} < 0,
\end{equation}
i.e. we have a {\em ferromagnetic}
interaction. Physically this is due to the fact 
that $\boldsymbol{S}_{1}$ and $\boldsymbol{S}_{5}$ are both coupled 
antiferromagnetically to $\boldsymbol{S}_{2}$; this makes it favorable to have
$\boldsymbol{S}_{1}$ and $\boldsymbol{S}_{5}$ parallel to each other, thus 
an effective ferromagnetic bond is generated. Cluster 2356, given by the 
Hamiltonian $H_3$, looks almost the same as cluster 1234 except that spin $\boldsymbol{S}_{5}$ and $\boldsymbol{S}_{6}$ are already connected by an original bond $J_{56}$. This original bond will be renormalized when $J_{23}$ is decimated together with $J_{25}$ and $J_{36}$. The renormalized bond is then given by :
\begin{equation}
\label{recursion3}
\tilde{J}_{56} = J_{56} + \frac{J_{25} J_{36}}{2 J_{23}}.
\end{equation}
The generated interaction between $\boldsymbol{S}_{5}$ and $\boldsymbol{S}_{6}$ is antiferromagnetic because they are sitting on the opposite sub-lattices. 

As discussed earlier, the effective F bonds generated by RSRG can lead to 
formations of effective spins with size larger than 1/2. We thus need to 
incorporate this possibility in our scheme, and
generalize the Hamiltonian in Eq.(\ref{h1}) and (\ref{h2}) by giving arbitrary sizes to the spin operators in the Hamiltonian and by having either sign for the couplings. We treat $H_I$ as a perturbation to $H_0$ as before. In the space of degenerate ground states of $H_0$, the spins $\boldsymbol{S}_1$ and $\boldsymbol{S}_2$ form a state of maximum total spin $S = S_2 + S_3$ for ferromagnetic ($J_2 < 0$) or of minimum total $S = |S_2 - S_3|$ for antiferromagnetic ($J_2 > 0$) while the spins $\boldsymbol{S}_1$ and $\boldsymbol{S}_4$ can point in any direction. The degenerate ground states span the Hilbert space $\mathcal{H}$ which is the product space of the spin spaces $\boldsymbol{S}_1$, $\boldsymbol{S}$, and $\boldsymbol{S}_4$. $H_I$ will partially lift the degeneracy in $\mathcal{H}$ and induce an effective Hamiltonian in $\mathcal{H}$. The effective Hamiltonian can be calculated using the projection theorem \cite{sakurai}:
\begin{equation}
H_{eff} = P H P,
\end{equation}
where $P$ is the projection operator that projects the full Hamiltonian $H$ into the subspace where $S$ is maximum(minimum). The detail of this calculation is
available in Ref. \onlinecite{westerberg}.
Here we just give the final result. After the strong bond is decimated, 
we can write down the effective Hamiltonian $H_{eff}$ as :
\begin{equation}
H_{eff} = \tilde{J}_{1} \boldsymbol{S}_{1}\cdot\ \boldsymbol{S} + \tilde{J}_{3} \boldsymbol{S}\cdot\boldsymbol{S}_{4} + constant,
\end{equation}
where
\begin{eqnarray}
\label{renorm2}
\tilde{J}_{1} &=& \frac{S(S+1)+S_2(S_2+1)-S_3(S_3+1)}{2 S(S+1)} J_1,\nonumber\\
\tilde{J}_{3} &=& \frac{S(S+1)+S_3(S_3+1)-S_2(S_2+1)}{2 S(S+1)} J_3,
\end{eqnarray}
where $S = |S_2 \pm S_3|$ depending on the sign of $J_2$. 

In the case where $J_2 > 0$ and $S_2 = S_3$, the ground state of the strong bond
is a singlet and there is no effective spin left after decimation. 
Second-order perturbation expansion yields a non-zero interaction between 
$S_1$ and $S_4$:
\begin{equation}
\tilde{J} = \frac{2}{3} S_2(S_2+1) \frac{J_1 J_3}{J_2}.
\end{equation}
It can be shown that the cases discussed above exhaust all possible situations
we may encounter when applying the RSRG to a spin ladder.

In implementing the RSRG procedure outlined above, one finds that each spin is
coupled to more other spins as more and more spins are decimated, and the
couplings can be either F or AF.
There is, however, one major simplification due to the
bipartite nature of the original lattice, which is also of physical importance
as we discuss below. 
In the 
beginning we have a lattice structure which can be divided into two sublattices
(A and B) in which spins in sublattice A get coupled only to spins in 
sublattice B. As we run our RG procedure this is no longer true. Not only 
spins from the same sublattice can get coupled together but also the sizes of
the spins are no longer the same. It becomes a relevant question to ask where
to put an effective spins formed by two spins with different sizes and what
the types of interactions are between this effective spin with the rest of the
lattice. We apply the majority rule in our RG scheme to incorporate this 
situation. The idea of this rule is to put the effective spin formed by two 
spins with different sizes connected by AF/F coupling on the sublattice where
the larger spin is. Using this rule we are able to show that two spins
sitting on opposite sublattices will always have AF couplings while those 
sitting on the same sublattice will always have F couplings. This is clearly
true in the beginning; we will show below the RG procedure combined with the
majority rule preserves this structure. Physically this simply reflects the
fact that the nearest neighbor AF couplings on a bipartite lattice   
has no frustration; they prefer the spins in the same sublattice to be 
parallel, and in opposite sublattices to be antiparallel.

Let us elaborate this idea in more detail to better understand the majority
rule. We 
have seen in our discussion above that there are three different cases which
exhaust all the possible combinations encountered in our RG procedure.
First, we have two spins with the same size 
connected by AF coupling. Second, two spins with the same or different sizes 
connected by F coupling and third, two spins with different sizes connected 
by AF coupling. These three cases are shown in Fig. \ref{fig:rg}b and c. 
In the first case we do not have to worry about applying the majority rule
because there is no effective spin formed. The configuration is shown on as
cluster 1234 on Fig.
\ref{fig:rg}b. We just use the recursion relation derived in Eq. 
(\ref{recursion}) to determine the type of interactions between the spins which
were the third nearest neighbors. Some of the possible sublattice combinations for this case is shown in Table (\ref{tab:sign}). Here it is clearly shown that
two spins sitting on opposite sublattices will have AF interactions and those 
sitting on the same sublattice have F interactions. 

\begin{table}
\begin{tabular}{c c c c c c c c}
1 & 2 & 3 & 4 & $J_{23}$ & $J_{12}$ & $J_{34}$ & $\tilde{J}_{14}$\\
\hline
A & A & B & A &     +    &     -    &     +    &         -       \\
A & A & B & B &     +    &     -    &     -    &         +       \\
B & A & B & A &     +    &     +    &     +    &         -       \\
B & A & B & B &     +    &     +    &     -    &         +       \\
\end{tabular}
\caption{Some possible sublattice combinations for $S_2 = S_3$ and $J_{23} > 0$}
\label{tab:sign}
\end{table}

The configuration for the second case is shown on Fig. \ref{fig:rg}c. 
We have already seen from
Table (\ref{tab:sign}) that for two spins to have a ferromagnetic coupling, 
they must be sitting on the same sublattice. 
In this case there is no ambiguity where to 
put the effective spin. We can choose the effective spin to be located 
on the site where either $\boldsymbol{S}_2$ or $\boldsymbol{S}_3$ is used to 
be located. We can figure out the sign
of the renormalized couplings in the same way as it is done in Table 
(\ref{tab:sign}). The renormalized coupling is given by :
\begin{equation}
\tilde{J}_{12} = \frac{S_2}{S_2+S_3} J_{12}.
\end{equation}
With this recursion relation and majority rule, we can determine the sign of 
the renormalized
couplings for all combinations possible. This is shown in Table 
(\ref{tab:sign2}). The conclusion that AF coupling is always on opposite 
sublattices and F coupling is always on the same sublattice remains valid. 

\begin{table}
\begin{tabular}{c c c c c c c}
1 & 2 & 3 & $S_{eff}$ & $J_{23}$ & $J_{12}$ & $\tilde{J}_{12}$ \\
\hline
A & B & B &     B     &     -    &     +    &        +         \\ 
B & B & B &     B     &     -    &     -    &        -         \\
\end{tabular}
\caption{Possible sublattice combinations for and $J_{23} < 0$.
The column $S_{eff}$ gives us the sublattice where we put the effective spin.}
\label{tab:sign2}
\end{table}

The last case is when $S_2 \ne S_3$ and $J_{23} > 0$. The majority rule tells
us to put the effective spin on the sublattice of the spin
with bigger size. If $S_2 > S_3$, we put the effective spin on the sublattice 
in which $\boldsymbol{S}_2$ is sitting and vice versa. The recursion 
relations for the couplings are given by the equation :
\begin{eqnarray}
\tilde{J}_{12} = J_{12}\frac{S_2+1}{S_2-S_3+1}.
\end{eqnarray}
The type of interaction between the effective spin and the rest of the lattice
is shown in Table (\ref{tab:sign3}), where we take an example $S_2 > S_3$.
The result is the same as the two previous cases where  AF coupling is always 
on opposite sublattices and F coupling is always on the same sublattice. 
Should we change $S_2 < S_3$, the result would remain valid. Table 
(\ref{tab:sign3}) shows the configurations where $S_2 > S_3$.

\begin{table}
\begin{tabular}{c c c c c c c}
1 & 2 & 3 & $S_{eff}$ & $J_{23}$ & $J_{12}$ & $\tilde{J}_{12}$ \\
\hline
A & A & B &     A     &     +    &     -    &        -         \\ 
B & A & B &     A     &     +    &     +    &        +         \\
\end{tabular}
\caption{Possible sublattice combinations for $S_2 > S_3$ and $J_{23} < 0$.
$S_{eff}$ gives us the sublattice where we put the effective spin.}
\label{tab:sign3}
\end{table}

We have thus shown that the application of the majority rule will preserve
the type of interactions between spins sitting on opposite sublattices or
the same sublattice. If the spins are sitting on the opposite sublattices,
the interaction is always antiferromagnetic and if on the same sublattice, the
interaction is always ferromagnetic. This conclusion can be generalized to 
higher dimensions as long as the original AF interactions couple only spins 
sitting on opposite sublattices. 

\section{Numerical Results}

We have carried out the renormalization scheme for the ladder as described in
section II numerically, with length of the ladders up to 20,000. In the 
decimation process, we pick up the strongest bond as defined by the absolute 
value of the bond strength,\cite{note}
decimate it, and calculate the renormalized couplings to the neighboring 
spins. This procedure is iterated until the number of spins in the ladder is 
about 3\% of the original number of spins. The initial distributions are taken 
to be in power-law form :
\begin{eqnarray}
\label{distribution}
P_{\parallel}(J_{i,j}) &=& (1-\alpha) J_{i,j}^{-\alpha}, 0 < J_{i,j} < 1;
\nonumber\\
P_{\perp}(K_i) &=& \frac{1-\alpha}{\Lambda^{1-\alpha}} K_{i}^{-\alpha},
0 < K_i < \Lambda.
\end{eqnarray}
Here $0 \leq \alpha < 1$ is the measure of disorder (the bigger $\alpha$, the
stronger the randomness strength), and $0 < \Lambda \leq 1$ is the anisotropy 
parameter; in the limit $\Lambda\rightarrow 0$ the two chains decouple. 
We use a power-law form for our initial distributions because in the case of
random spin chains, fixed point distributions at low energies typically have
a power law form; we can thus hope to be able to approach the low-energy
fixed points faster by starting with a power law distribution. 

\begin{figure*}
\includegraphics*[scale=1.2]{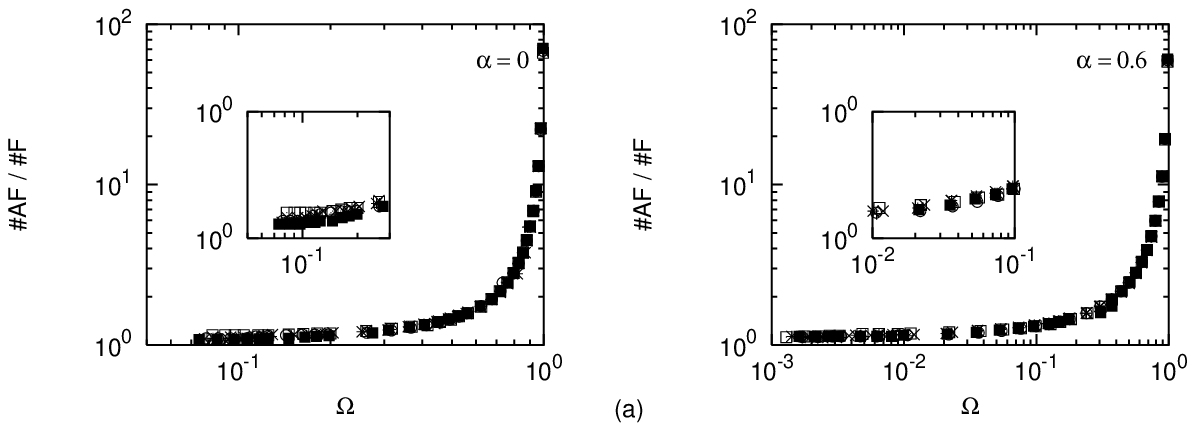}
\includegraphics*[scale=1.2]{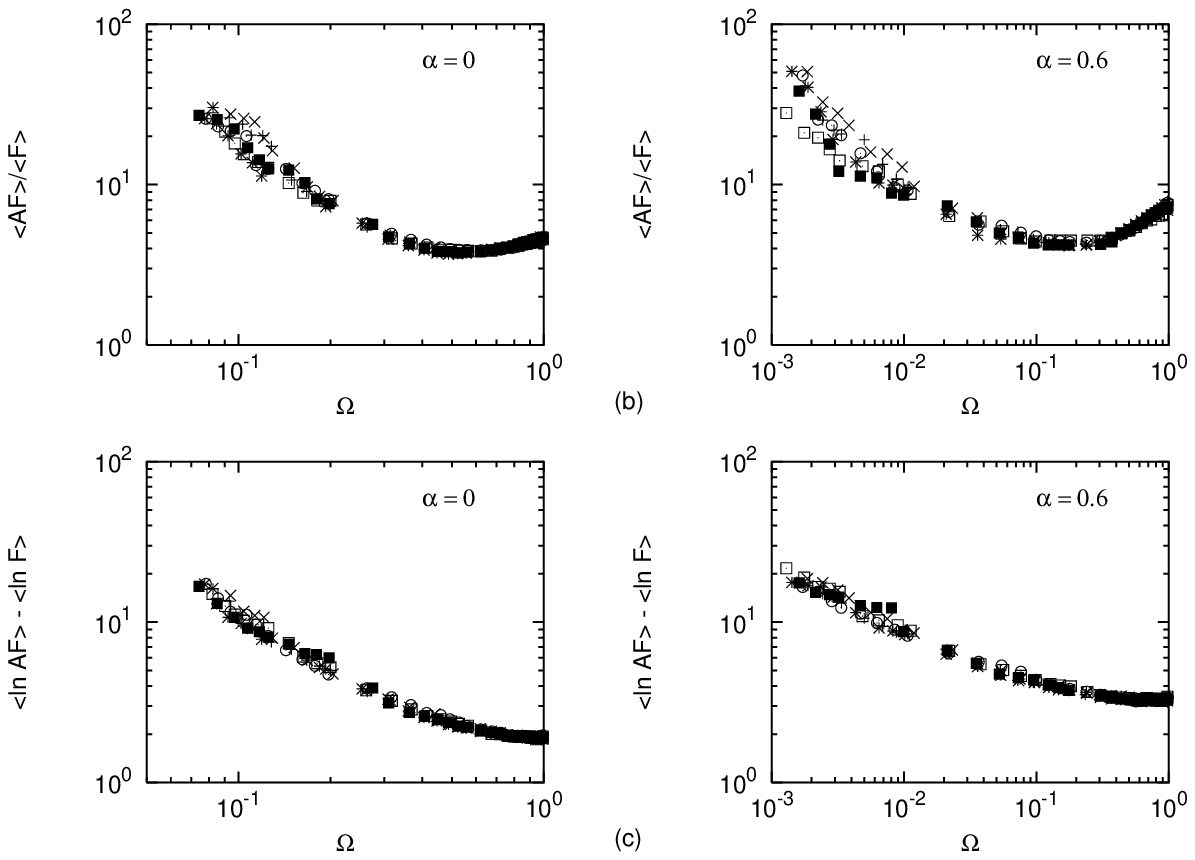}
\includegraphics*[scale=1.2]{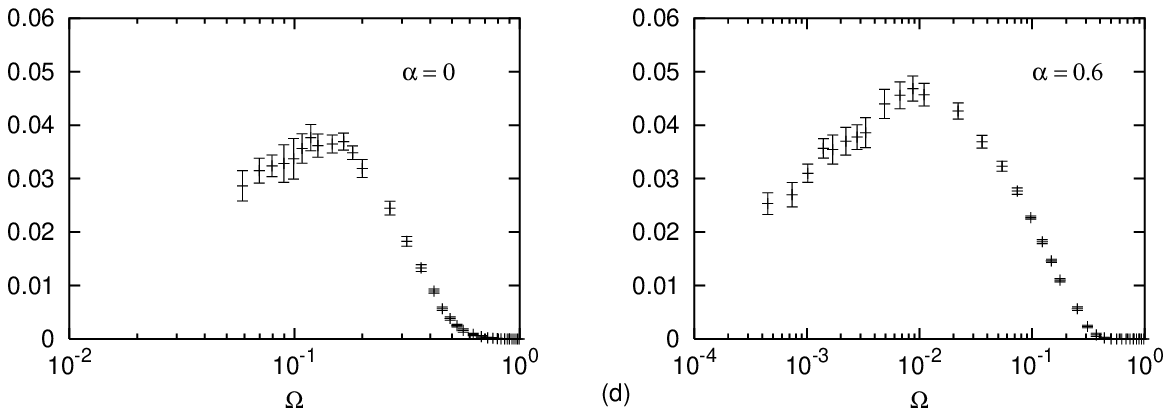}
\caption{Numerical simulation results of the proliferation of ferromagnetic 
bonds and larger spins, for two different initial bond distribution
($\Lambda=1$ in both cases). 
(a) The ratio of numbers of AF and F bonds; (b) the ratio of average strength
of AF and F bonds; (c) the difference of the averaged logarithm of AF and F 
bond strengths, 
and (d) the ratio of number of spins larger than 1/2 and the total spins, all
as functions of energy scale $\Omega$. All samples used have size $N=20,000$ and
in (a)-(c) different symbols represent different samples.}
\label{fig:ratio}
\end{figure*}

As discussed earlier, due to the presence of F bonds generated by RSRG, 
effective spins with size bigger than 1/2 appear at low energies. One might
think that such larger spins may proliferate, and the typical size of the spins
may grow indefinitely, leading to to a phase dominated by weakly coupled large
spins. This was found to be the case in spin chains with random AF and F 
couplings studied by Westerberg et al.\cite{westerberg} We find, however,
this is {\em not} the case in the present problem. 
We address the issue of proliferation of F bonds and large spins in 
Fig. \ref{fig:ratio}, where data for $\alpha=0$ and $\alpha = 0.6$ 
(both with $\Lambda=1$) are shown. We plot the ratio of the numbers of AF bonds 
and F bonds as a function of bond strength cutoff $\Omega$ in
(a). At the early stages of RG the system
consists of a large fraction of AF bonds and a small percentage of F bonds 
generated by the decimation process. As the energy scale is lowered more F 
bonds are generated and more AF bonds are removed so the ratio of the number 
goes down. In the low energy limit, we find the number of F 
bonds is very close to the number of AF bonds. This can be seen more clearly at 
the insets in Fig. \ref{fig:ratio}a. 
Even though the numbers of AF and F bonds are almost 
equal, the strengths of AF and F bonds behave completely differently in this 
limit. AF bonds always dominate the system. In (b) we plot the ratio of the 
average strength of AF and F bonds. When the bond cutoff $\Omega$ goes below
0.2, where the numbers of AF and F bonds are almost equal, the ratio of the 
averages grows rapidly which means the AF bonds are much stronger than the 
F bonds in the low energy limit. In (c) we plot the difference of the averaged
logarithms 
of AF and F bonds; the exponential of this quantity reflects the ratio between
{\em typical} AF and F bonds. Similar to (b), here we see the the difference
grows very fast, again showing the dominance of AF bonds over the F bonds. 
In (d) we plot the sample averaged ratio of the number of spins larger than 1/2
to the total number of spins. Here we see that while larger spins do appear, 
their percentage remains small, and the percentage {\em decreases} with
the cutoff $\Omega$ going down in the low-energy limit. Another piece of 
information that is not included in the figure is that most of the larger spins 
are spin-1's, with a very small percentage of spin 3/2 and spin 2. We have not 
found any trace of spins larger than 2 in our simulations. 
We find qualitatively similar behavior in all initial distributions we have 
looked at, indicating this is generic.

Physically, such behavior has its origin in the bipartite nature of the lattice
structure of the 2-leg ladder. As we have shown earlier, the effective couplings
generated by the RSRG is always AF between spins of opposite sublattices, and 
F between spins of the same sublattice. Since the numbers of spins in the two
sublattices are the same, the number of F and AF bonds become very close in
the low-energy limit. On the other hand spins in opposite sublattices tend to
be closer to each other, leading to the fact that AF bonds dominate F bonds in
strength. This in turn suppresses formation of large spins.

\begin{figure*}
\includegraphics*[scale=1.2]{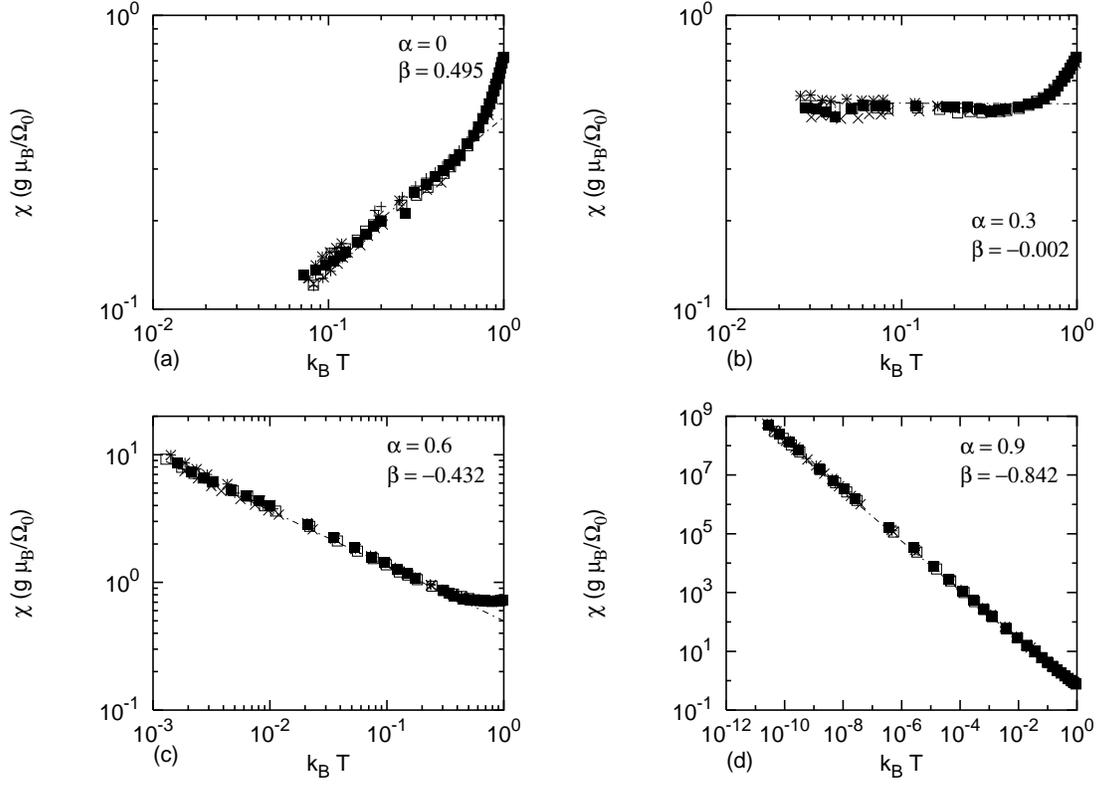}
\caption{The susceptibility per spin as a function of temperature for several 
disorder strengths $\alpha$. The anisotropy parameter $\Lambda = 1$ is fixed,
and the system size is $N=20,000$. 
The power law exponents, $\beta$, are calculated using the least square fit to
the low-temperature data. The different symbols in the figure correspond to 
different samples. We do not take the sample average because the sample to 
sample variations are small.} 
\label{fig:sus}
\end{figure*}

Our most important results are presented in Figs. 3-6, where we plot the 
temperature dependence of the spin susceptibility, and the ground state
spin-spin correlation function.
The susceptibility is calculated as the following. We proceed with the RSRG 
until the bond cutoff $\Omega$ is equal to the
temperature $T$. We neglect contributions of spins that have already been 
decimated, and treat the remaining spins as free spins, thus
their contribution to the susceptibility is just the Curie susceptibility. This
is a good approximation as long as the bond distribution is broad.
The total susceptibility is thus given by :
\begin{equation}
\chi_{tot} = \frac{g \mu_B}{3k_B T} \sum_s N_s s(s+1),
\end{equation}
where $N_s$ is the number of spins left at energy scale $\Omega=T$ for a given 
spin size $s$ and the summation runs over all possible spin sizes.
In Fig. \ref{fig:sus} we plot the susceptibility per 
spin for different samples as a function of temperature for different disorder 
strength $\alpha$, all with isotropic coupling ($\Lambda=1$). 
In all cases we find the low-$T$ susceptibilities can be fit quite well to 
power-law
dependence on $T$: $\chi\sim T^{-\beta}$; 
the power-law exponent $\beta$, which we obtain from a 
least-square fit to the low-$T$ part of the data, is non-universal; it can 
describe both 
divergent ($\beta > 0$) $\chi$
for stronger randomness (larger $\alpha$), or vanishing $\chi$
($\beta < 0$) for weaker randomness (smaller $\alpha$), as $T\rightarrow 0$. 
It is worth
noting that for $\Lambda=1$, we always have $\beta < \alpha$, and such behavior
persists for very strong disorder like $\alpha=0.9$.
Such behavior is very 
different from random AF spin-1/2 chain with any amount of randomness, or
random AF spin-1 chain with sufficiently strong randomness,
where the system flows to the so-called random singlet (RS) fixed point, in
which the bond distribution is {\em infinitely broad}, 
the spin-spin correlation follows
a universal power-law, and the
susceptibility diverges in a universal manner\cite{fisher1}:
\begin{equation}
\chi \sim 1/(T \ln^2(\Omega_0/T)).
\label{rs}
\end{equation}
Instead, the fact that we find power-law exponent $\beta$ to be always less
than 1 indicate the width of the bond distribution is {\em finite}.
Of course, in principle we can not completely rule out the possibility that our
system size (and correspondingly, temperature range) is not wide enough for
us to approach the true low-$T$ asymptotic behavior of $\chi$, which for 
strong enough randomness may be controlled by a fixed point with infinitely broad
bond distribution and universal. We believe, however, this is highly unlikely
for the following reasons. (i) Our power-law fit already extends to a very wide
range in $T$. In particular, for $\alpha=0.9$, a single power-law fits all
the data very well that is over eleven orders of magnitudes in $T$, 
with no indication of crossover
to other behavior at low $T$. (ii) As we will see later, the spin-spin 
correlation function appears to be short-ranged, indicating that the 
long-distance, low-temperature physics is {\em not} controlled by a single 
scale-invariant fixed point.

\begin{figure*}
\includegraphics*[scale=1.2]{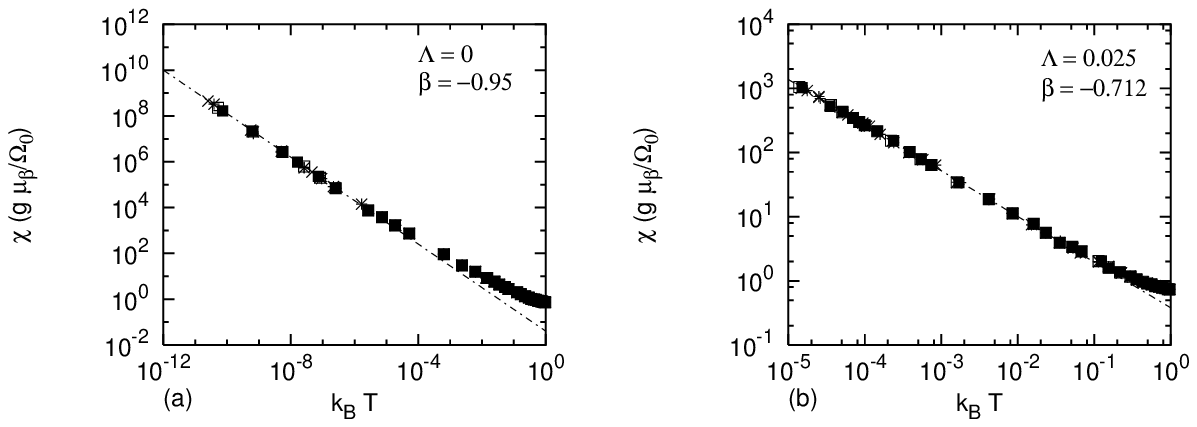}
\includegraphics*[scale=1.2]{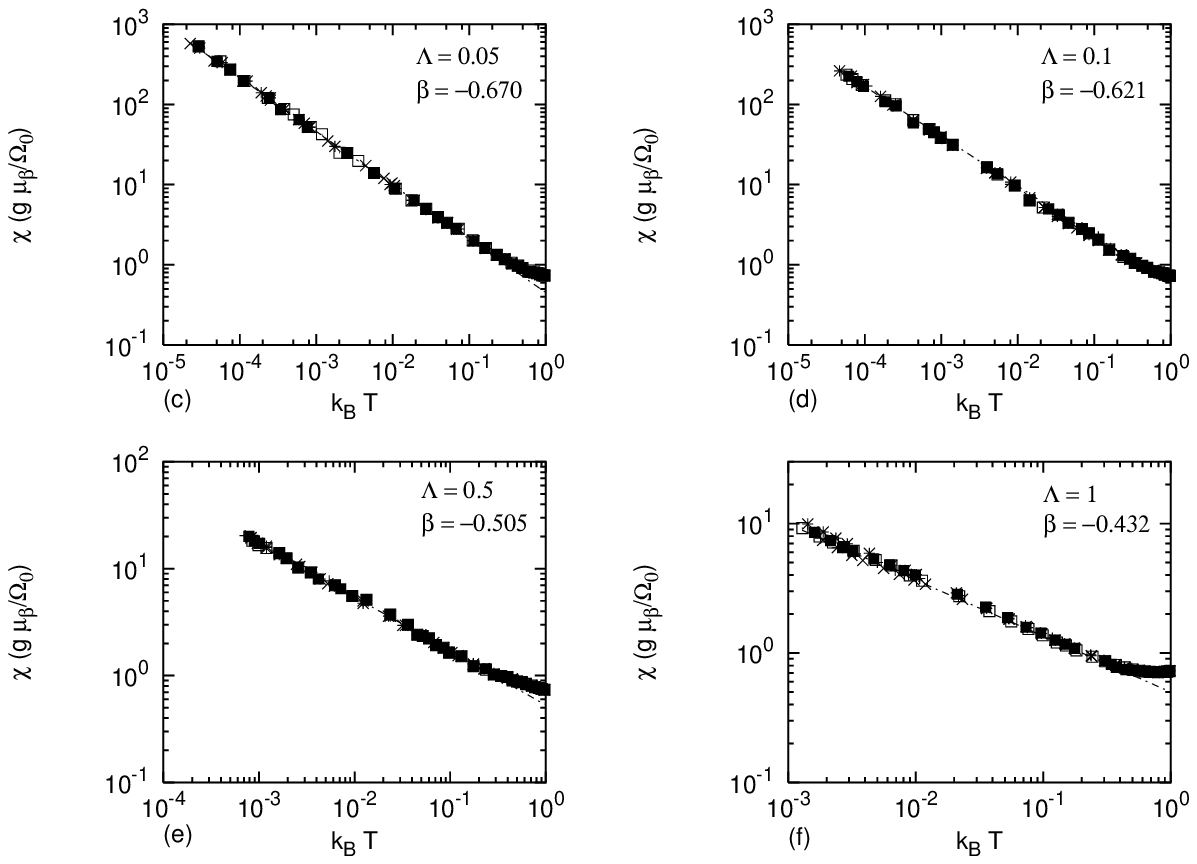}
\caption{The susceptibility per spin as a function of temperature for a 
given $\alpha = 0.6$ with varying anisotropy parameter $\Lambda$. System size
fixed to be $N=20,000$.
The power law exponent increases as we decrease the anisotropy parameter. For $\Lambda = 0$ we have decoupled chains. We plot all different samples in the figure without taking the sample average because the variations are very small.}
\label{fig:sus2}
\end{figure*}

In the absence of interchain coupling, the ladder becomes two decoupled random
AF spin-1/2 chains, where the long-distance, low-temperature physics {\em is}
controlled by the RS fixed point and universal.
To address how the system crosses over from one behavior to another we have
studied how the susceptibility varies with the anisotropy parameter $\Lambda$.
In Fig. \ref{fig:sus2} the susceptibility per spin for different values of 
$\Lambda$ is presented, for a fixed $\alpha=0.6$.  
Again, we find non-universal behavior here.
As we vary $\Lambda$ from 1 to 0, the power-law exponent of the susceptibility 
increases continuously. 
In the case of $\Lambda = 0$ we have decoupled chains and the susceptibility 
is expected to follow Eq. (\ref{rs}). While for a finite range of $T$ it can
be fit reasonably well to a power-law with $\beta$ very close to 1, the small
upward curvature of the data indicates $\beta$ would increase as one goes to 
lower $T$, consistent with Eq. (\ref{rs}). On the other hand a very weak 
interchain coupling ({\em e.g.}, $\Lambda=0.025$) leads to a significant change
it $\beta$, and there is no longer obvious upward curvature in the data. This
suggests that a {\em weak} interchain coupling immediately destabilizes the
RS fixed point.   

\begin{figure*}
\includegraphics*[scale=1. ]{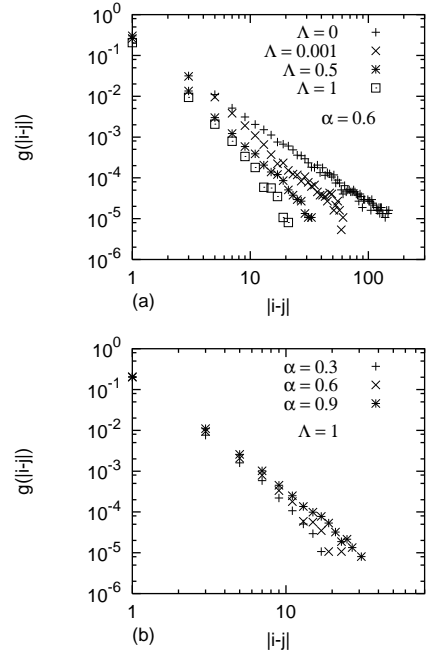}
\caption{The sample-averaged
spin-spin correlation along the chain (a) with varying $\Lambda$ and 
$\alpha = 0.6$, (b) with varying $\alpha$ and $\Lambda = 1$, for N = 20,000. 
The fit to power-law behavior yields the following power-law exponents : 
(a) $\nu = 1.97$ for $\Lambda = 0$, $\nu = 2.56$ for $\Lambda = 0.001$, 
$\nu = 2.87$ for $\Lambda = 0.5$, $\nu = 3.23$ for $\Lambda = 1$, and 
(b) $\nu = 3.45$ for $\alpha = 0.3$, $\nu = 3.23$ for $\alpha = 0.6$, 
$\nu = 2.89$ for $\Lambda = 0.9$.}
\label{fig:corr}
\end{figure*}

We now turn the discussion to the ground state
spin-spin correlation function along the
chain:
\begin{equation}
g(|i-j|)=(-1)^{i-j}<<\boldsymbol{S}_{i,k}\cdot\boldsymbol{S}_{j,k}>>,
\end{equation}
where $<< >>$ stands for both quantum and disorder average.
We calculate $g(|i-j|)$
in the following way. 
We run the RSRG until all spins are decimated, 
and then simply count the number
 of singlet pairs formed for a given distance $|i-j|$, divide this number by 
the total number of pairs and multiply the result by 3/4. In the RS phase,
$g(|i-j|)\sim |i-j|^{-2}$.\cite{fisher1} In Fig. \ref{fig:corr}a we study how the 
interchain interaction affects the correlation along the chain by varying the 
anisotropy parameter $\Lambda$. 
Fitting the data to a power-law dependence: $g(|i-j|)\sim |i-j|^{-\nu}$,
we obtain $\nu = 1.97$ for $\Lambda=0$ (decoupled chain case), 
which is very close to the analytical
result $\nu=2$.\cite{fisher1} For nonzero $\Lambda$, the 
correlation decays much faster than that of the chain. Even a small amount of 
interchain interactions (say, $\Lambda = 0.001$) change the behavior of the 
correlation considerably. 
We can see a downward curvature 
in the data, which is particularly obvious for $\Lambda = 1$ and 0.5,
indicating the short
range (decaying faster than any power-law) behavior of the correlation. 
If we try to fit the ground state 
correlation for non-zero $\Lambda$ to a power-law, we would get 
considerably larger 
power-law exponent $\nu$, even for $\Lambda$ as small as 0.001.
This strongly suggests that introduction of interchain 
interactions immediately destabilizes the RS phase that controls the 
low-energy of the decoupled chains, and leads to short-range spin-spin
correlation in the ground state. 
We have also calculated how the correlation 
changes as we vary $\alpha$ for fixed $\Lambda = 1$,
in Fig. \ref{fig:corr}b. Here we find while stronger randomness (larger $\alpha$)
tends to enhance correlation at large distances, the correlation is still
short-ranged for very strong randomness ($\alpha=0.9$) as evidenced by the 
downward curvature of the data. 

\begin{figure*}
\includegraphics*[scale=1. ]{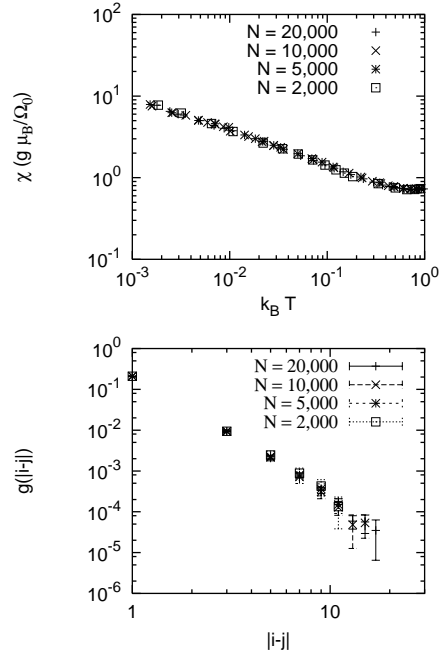}
\caption{(a) The susceptibility per spin as a function of temperature, 
(b) the spin-spin correlation along the chain, for $\alpha = 0.6$ and 
$\Lambda = 1$ with varying system sizes. 
No significant variation in these two quantities as the system sizes are varied. Both (a) and (b) are sample averages.}
\label{fig:finite}
\end{figure*}

One general concern in numerical calculations of the kind discussed here is
finite-size effects. We show in Fig. \ref{fig:finite} that the system sizes 
we use in this work are large enough that
the finite-size effects are negligible. 
The sample averages of the susceptibility per spin do not 
show any noticeable fluctuations as the system sizes are varied from $N=2,000$
to $N=20,000$. The same is true
for the ground state spin-spin correlation. At large separation there are some 
variations due to sample to sample fluctuations. 
We can thus
safely 
conclude that the finite-size effect is negligible in our study. 

\section{summary and discussions}

In this work we have used the RSRG method to study an AF two-leg spin-1/2 
ladder, with strong bond randomness. We find that the spin susceptibility is
non-universal, and the ground state spin-spin correlation is short-ranged, for
any randomness and interchain coupling strength. For sufficiently strong
randomness or sufficiently weak interchain coupling, the spin susceptibility
exhibit power law divergence as $T\rightarrow 0$, which is characteristic of
quantum Griffith behavior.

Melin {\em et al.}\cite{melin} used the RSRG method as well as density matrix
RG to study the distribution of the energy gap separating the ground and first
excited states in clusters (with length up to 512)
of AF two-leg spin-1/2 ladders.
They find that the dynamic exponent $z$ that characterizes this distribution is
non universal and depends continuously on randomness strength. Based on this they
conclude that the low-energy physics of the system is controlled by a fixed 
point with a finite width in the bond distribution function, and the system is
in a quantum Griffith phase. Our results and conclusions agree with theirs.

It is by now well established that in the absence randomness, the two-leg 
AF spin-1/2 ladder supports a finite excitation gap, and the spin-spin
correlation is short-ranged. It is generally true that randomness tends to 
introduce low-energy excitations, which can lead to divergent spin 
susceptibilities as found here. Our results indicate however, despite the 
low-energy excitations introduced, the {\em phase} with short range spin-spin
correlation appears to be stable against any amount of randomness. This is 
certainly consistent with Ref. \onlinecite{og}, where the authors find the 
pure ladder to be remarkably stable against various kinds of disorder. On the
other hand this is very different from the AF spin-1 chain, where sufficiently
strong bond randomness drives the system from the Haldane phase to the random
singlet phase with universal thermodynamics and power-law spin-spin 
correlation, through a second-order phase transition.\cite{hy,monthus}

As discussed earlier, for the present system the bond distribution has a 
{\em finite} width in the low-energy limit, no matter how strong the randomness
is initially. This indicates that the RSRG method is {\em not} asymptotically
exact when applied to the present model. However this method is quantitatively 
accurate as long as the randomness is strong, we thus believe the qualitative
conclusions we draw from our results are robust. 

\acknowledgments
This work was supported by NSF grant No. DMR-9971541, and the Research 
Corporation. K.Y. was also supported
in part by the A. P. Sloan Foundation.

\end{document}